# Automatic Assessment of Dysarthria Using Audio-visual Vowel Graph Attention Network

Xiaokang Liu, Xiaoxia Du, Juan Liu, Rongfeng Su, Manwa Lawrence Ng, Yumei Zhang, Yudong Yang, Shaofeng Zhao, Lan Wang, *Member, IEEE* , Nan Yan, *Member, IEEE*

*Abstract--***Automatic assessment of dysarthria remains a highly challenging task due to high variability in acoustic signals and the limited data. Currently, research on the automatic assessment of dysarthria primarily focuses on two approaches: one that utilizes expert features combined with machine learning, and the other that employs data-driven deep learning methods to extract representations. Research has demonstrated that expert features are effective in representing pathological characteristics, while deep learning methods excel at uncovering latent features. Therefore, integrating the advantages of expert features and deep learning to construct a neural network architecture based on expert knowledge may be beneficial for interpretability and assessment performance. In this context, the present paper proposes a vowel graph attention network based on audio-visual information, which effectively integrates the strengths of expert knowledges and deep learning. Firstly, various features were combined as inputs, including knowledge based acoustical features and deep learning based pre-trained representations. Secondly, the graph network structure based on vowel space theory was designed, allowing for a deep exploration of spatial correlations among vowels. Finally, visual information was incorporated into the model to further enhance its robustness and generalizability. The method exhibited superior performance in regression experiments targeting Frenchay scores compared to existing approaches.**

*Index terms*—Dysarthria Assessment, Vowel Graph, Graph Attention Network.

Xiaokang Liu, Xiaoxia Du and Juan Liu contribute equally to the article.
Xiaokang Liu and Juan Liu are with ICAS Key Laboratory of Human-Machine Intelligence-Synergy Systems, Shenzhen Institute of AdvancedTechnology, Chinese Academy of Sciences, Shenzhen, 518055, China. University of Chinese Academy of Sciences, Shenzhen, 518055, China.
Xiaoxia Du is with Department of Neurorehabilitation, Beijing Boai Hospital, China Rehabilitation Research Center, Beijing 100068, China.
Manwa Lawrence Ng is with Division of Speech and Hearing Sciences, University of Hong Kong, Hong Kong 999077, China.
Yumei Zhang Department of Rehabilitation Medicine, Beijing Tiantan Hospital, Capital Medical University, Beijing 100070, China.
Rongfeng Su and Yudong Yang are with Shenzhen Institutes of Advanced Technology, Chinese Academy of Sciences, Shenzhen 518055, China.
Shaofeng Zhao is with Department of Rehabilitation Medicine, The Eighth Affiliated Hospital of Sun Yat-sen University, Shenzhen 518055, China (e-mail: zhaosf1@163.com).
Lan Wang and Nan Yan are with Guangdong-Hong Kong-Macao Joint Laboratory of Human-Machine Intelligence-Synergy Systems , Shenzhen Institutes of Advanced Technology,Chinese Academy of Sciences , Shenzhen, 518055, China (e-mail: lan.wang@siat.ac.cn, nan.yan@siat.ac.cn)

## I. INTRODUCTION

DYSARTHRIA is a speech disorder that arises from damages to the motor neurons within speech-motor system, leading to various degrees and types of speech impairment in affected individuals. Being able to detect dysarthria holds practical significance in clinical applications, and accurate assessment of dysarthria is a prerequisite for formulating effective and efficient rehabilitation plans. Thanks to its comprehensiveness, the Frenchay Dysarthria Assessment (FDA) has become the gold standard for clinical diagnosis of motor-type dysarthria [1], [2]. However, the subjective assessment included in the FDA heavily relies on the experience of doctors or speech experts, and diagnostic outcomes can be greatly influenced by various factors such as familiarity of the doctor with the patient, speech context, and the impact of semantic and syntactic features. Furthermore, the FDA assessment is complex, requiring measurement of indicators across eight categories totaling 28 dimensions [3], resulting in subjective assessments using the FDA expensive and time-consuming [4]. In comparison to subjective assessment methods, objective assessment methods for the FDA offer advantages such as its repeatability, low cost, and high reliability. These facilitate early screening of dysarthria and remote rehabilitation monitoring [5]. Assessment of dysarthria has been extensively investigated and results are available in the literature [6-21]. However, current solutions can only categorize the overall severity of dysarthria, lacking finer severity rating such as scoring, and are incapable of effectively evaluating specific sub-items of dysarthria. For healthcare professionals, a more detailed assessment of dysarthria including that of sub-items holds greater clinical application.

Thus, accurately conducting objective assessments using the FDA still remains challenging. First, the etiology of dysarthria is complex, and various neural injuries responsible for the articulatory system may lead to the emergence of dysarthric speech. The wide variety of dysarthria etiology implies high heterogeneity and data scarcity. Precision in assessing dysarthria requires more targeted speech features. Conventional speech signal features (such as MFCC, fundamental frequency, resonance peaks, etc.) may not be useful with dysarthria speech due to the impaired pronunciation and articulatory organ dysfunction [22], resulting in insufficient representation capabilities of conventional speech signal features and models for dysarthria speech. Secondly, the manifestations of dysarthric speech are diverse, impacting the entire articulation process. This explains the need for multiple observation methods in order to yield a comprehensive assessment. For instance, in the FDA, not only speech information but also a substantial amount of visual information is required, encompassing evaluations of various sub-items (including reflexes, respiration, lips, jaw, palate, larynx, tongue, and intelligibility) [3].

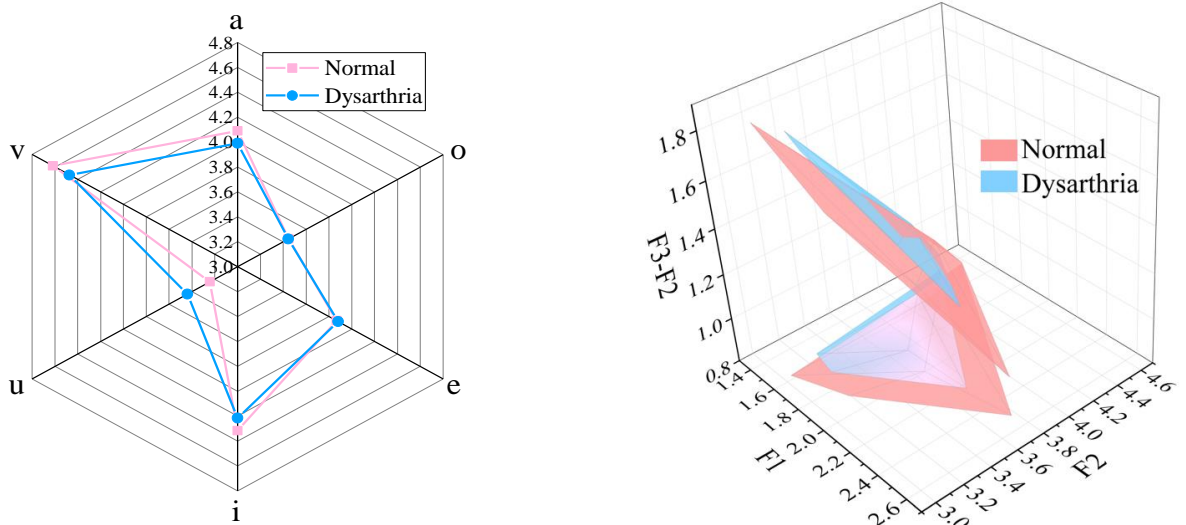

Fig. 1. Vowel space. The left panel shows a radar chart, while the right panel displays a three-dimensional graph. In the radar chart, the distribution of vowel space is presented using a scale measured in units of $\sqrt{F1^2 + F2^2}$ in bark. The three-dimensional graph depicts F1 on the X-axis, F2 on the Y-axis, and (F3-F2) in bark units on the Z-axis. The projection onto the X-Y plane represents a simplified view of this three-dimensional space. Notably, the vowel space of patients with dysarthria appears significantly reduced compared to that of normal individuals.

In addition, studies revealed distinct differences between dysarthria and normal speech with regard to the vowel space [23]. Discernment and resilience are observed between typical individuals and dysarthric patients in high-dimensional vowel space, as depicted in Fig 1. The vowel space for associated with normal individuals and dysarthric patients are also observed in the Previous Mandarin Dysarthria study [24]. These findings indicate that variances in vowel interrelation could represent a critical aspect in the diagnosis of dysarthric speech. Thus, a Vowel Graph Attention Network grounded in vowel space theory and attention mechanisms was devised in this paper to comprehensively capture vowel spatial information across different dimensions. Utilizing expert acoustical features of six cardinal vowels as input, the system applies attention mechanisms derived from vowel space theory to discern and incorporate vowel space relationships within the features. The procedure involves initial sample space construction based on vowel space theory to enhance the robustness of sample vowels' features and address data scarcity challenges. In addition, a bi-modal dysarthria assessment model was designed by leveraging the complementarity of audio and visual information for predicting FDA sub-item scores through regression.

In summary, the main contributions in this paper include the following: a method for constructing vowel space samples to augment data and implement dysarthria FDA regression assessment based on the Audio-visual Vowel Graph Attention Network (AV-VGAN) framework. The AV-VGAN framework involves two stages of fusion: one integrating phonetic space information with feature information and the other fusing audio and video information. The proposed framework underwent evaluation and compared to the state-of-the-art method, such as CNN, RESNET and WAV2VEC 2.0[25], [26]. Furthermore, we introduced a more interpretable set of audio-visual features for dysarthria. Ablation experiments confirmed the complementarity of audio and video information and the effectiveness of vowel space modeling.

This paper is structured as follows: the second section introduces related work, the third section outlines the proposed method, the fourth part introduces the database and experimental settings, the fifth part presents experimental results and analysis. The sixth part discusses the advantages of vowel features and explores aspects related to vowel space; the seventh part includes a list of references.

## II. RELATED WORK

Currently, research on the automatic assessment of post-stroke dysarthria based on speech information focuses primarily on two aspects. Firstly, there is an emphasis on interpretability, exploring features based on acoustic and physiological principles. Secondly, researchers are delving into examining different network structures through data-driven approaches. Given the perceptual differences between dysarthric and typical speech, an intuitive approach seems to involve analyzing speech features of individuals with dysarthria. As dysarthric speech is perceptually different compared to normal speech, analyzing the features present in the speech of individuals with dysarthria in order to train a classification machine is needed. With the accumulation of dysarthria data and the advancement of deep learning technologies, there is a growing interest in leveraging the powerful modeling capabilities of deep learning to unearth deeper-level features, making it a research hotspot.

### A. Feature-based assessment method for dysarthria

Current research exploring dysarthria features takes different approaches. Some studies analyzed these features from acoustic signals, including time domain, frequency domain, or phase of speech signals. Other studies explored the physiological features of dysarthric speech production, including features such as fundamental frequency stability, pronunciation stability, and resonance peak regularity. Common acoustic features are employed for the assessment of dysarthria. These include Mel-Frequency Cepstral Coefficients (MFCC) [27], multi-resolution MFCC [28], Perceptual Linear Prediction Cepstral Coefficients (PLPCCs) [29] and Spectral Frequency. Another direction was to explore effective features from the mechanics of pronunciation. These included auditory features related to fundamental frequency and resonance peaks [12][13], glottal features [16-18], and vowel-related features [10-12]. Physiologically, fundamental frequency or pitch is closely correlated with the rate of vibration of the vocal cords. Perturbation in fundamental frequency, known as jitter, results from uncontrolled vibrations of the vocal cords. Shimmer refers to variations in the amplitude of the speech signal, which can reflect the reduced tension and wear of the vocal cords. This is why jitter and shimmer were used for objective assessment of dysarthria in some studies [14]. Additionally, loudness, rate, and rhythm were also used to evaluate dysarthria [30] [31]. Vowels are a core component of syllables and have a significant impact on speech intelligibility. One indicator of dysarthria is whether vowels are distorted. These acoustic metrics have been used by speech and language pathologists (SLP) in the study of speech development [23], vowel recognition, and speaker features in dysarthria such as post-stroke dysarthria [32], cerebral palsy dysarthria [33], or Parkinson's disease (PD) dysarthria [34]. Features representing different kinds of information have been combined to enhance the performance of dysarthria assessment systems [23], [35], [36]. In a study by Belalcazar-Bolaños and colleagues [37], an expert feature set was used to evaluate dysarthria caused by Parkinson's disease. This feature set includes Harmonics-to-Noise Ratio (HNR), Normalized Noise Energy (NNE), Cepstral HNR (CHNR), and Glottal-to-Noise Excitation Ratio (GNE). Rong et al. [38] expressed interpretability as a linear weighted combination of speech, pronunciation, nasal, and rhythm features. Traditional amplitude spectrum features have demonstrated their importance in dysarthria speech processing.

### B. Assessment method of dysarthria based on deep learning

With the advancement of machine learning technologies, objective assessment models for dysarthria processing have been established. Yet, the limited scale of dysarthria data has restricted the use of large-parameter neural network models, such as Hidden Markov Models (HMM) using Maximum Likelihood Estimation (MLE) techniques [39], Gaussian Mixture Models (GMMs), and Long Short-Term Memory (LSTM) networks [7]. Using shallow classifiers alone may not fully exploit the advantages of machine learning in mining deep representations of specific aspects. Some studies therefore sought to increase network depth using various methods, such as Deep Belief Networks, multi-task learning, and short-time domain approaches[8], [9], [40]. Moreover, as speech recognition technologies progressed, some studies have utilized transfer learning with speech recognition models to address dysarthria assessment. Liu et al. [41] leveraged intelligibility information provided by an Automatic Speech Recognition (ASR) model to assess the severity of dysarthria. Their approach relied on phoneme posterior probabilities output by a deep neural network-based ASR system. These ASR-based speech features aimed to effectively quantify the mismatch between dysarthria speech and normal speech. The severity prediction was initially performed at the utterance level, and the individual utterance prediction scores were then aggregated for an overall assessment of the subject.

### C. Assessment method of dysarthria based on multi-mode

An increasing body of research suggests that, in situations where data is limited or in special scenarios, multimodal data tends to outperform unimodal data due to its ability to capture more information. For example, Ma et al. [42] proposed an audio-visual speech recognition model based on the Conformer structure, demonstrating superior speech recognition performance on multiple public datasets for the audio-visual multimodal compared to unimodal data. To objectively assess dysarthric speech, most methods relied on acoustic signals of speech and ignored the corresponding visual information. However, in a clinical setting, clinicians typically utilize multimodal data, usually both acoustic and visual modalities, in the assessment and diagnosis of dysarthric speech. The visual modality not only provides additional information in addition to acoustics, it also enhances the robustness of the system. For example, Area et al. [43] compared two different clarity measures obtained from the acoustic and audio-visual modalities for four speakers with dysarthria secondary to cerebral palsy (2 with mild-to-moderate dysarthria and 2 with severe dysarthria). Their results showed significant modality-dependent effects, with slightly higher scores in the audio-visual modality compared to the speech-only modality. Tong et al. [44] proposed a cross-modal deep learning framework for audio-visual input to classify the severity of dysarthria. Their results indicated that the cross-modal framework outperformed previous unimodal systems that made use of only audio data on the UASPEECH dataset.

## III. METHODS

### A. Sample space construction

The issue of limited yet highly variable data has been persistent challenges to efforts in enhancing the capabilities of dysarthria models[45]. In the present paper, a sample space construction method based on the vowel diagram is proposed to augment data and to reduce data variability. According to the theory of the acoustic vowel chart [46], the amplitudes of the first three formants of the six cardinal vowels can essentially predict the general range of movement of the tongue, lips, and other articulatory organs during pronunciation (see Fig 2). Inspired by this, the present project introduced a sample space construction method based on the six cardinal vowels of Mandarin Chinese that provided a comprehensive depict of patient's production. Simultaneously, it offers a more stable sample space to reduce data variability.

This method initially extracts all syllables from the sentences X, splitting words or sentences into syllables $S=\{s_1, s_2, s_3,…, s_T\}$ if necessary. Subsequently, it classifies all syllables according to the six vowels in Mandarin Chinese (a, o, e, i, u, v) and get $C =\{c_a, c_o, c_e, c_i, c_u, c_ü\}$. Finally, within each category, one sample is selected to form a set of six vowels, constituting a sample space $y_n$. The finally augmented data $Y_{all}$ is composed of multiple groups of $y_n$. The specific steps are described in algorithm 1.

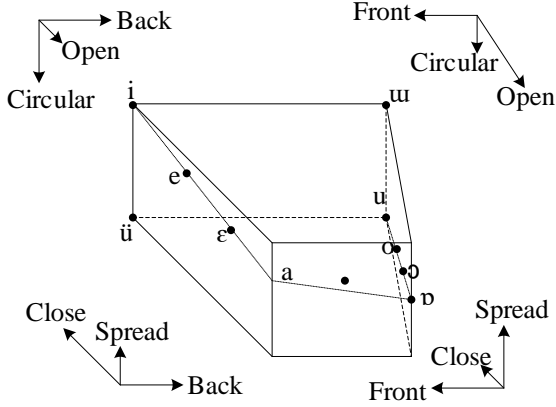

Fig. 2 A three dimensional model of the vowel space. Columns indicate vowel height (openness), with closed vowels placed at the top, while rows indicate vowel tongue positions (front-back locations), with front vowels positioned on the left side of the chart.

**Algorithm 1** Description of the data augmentation

**Inputs: $X_{all}$ all segment of a person**
1: for $X_n = X_1$ To $X_N$ do
2:     $S = \{s_1, s_2, s_3, \ldots, s_T\} = \text{split}(X_n)$
3:     for $s_t = s_1$ To $s_T$ do
4:         for $v$ in $\{v_a, v_o, v_e, v_i, v_u, v_ü\}$ do
5:             if $v$ in $s_t$ then
6:                 $C_v$ add $s_t$
7:             end if
8:         end for
9:     end for
10: end for
11: for $y_n$ in $\text{zip}(C_a, C_o, C_e, C_i, C_u, C_ü)$ do
12:     $Y_{all}$ add $y_n$
13: end for
**Output: augmented data $Y_{all}$**

### B. Feature extraction

Evidence suggests that dysarthria significantly affects both acoustic and visual characteristics, encompassing phonation, articulation, prosody, intelligibility, and lip movements [3]. However, current dysarthria features often focus on specific aspects, overlooking its overall nature. The present study aimed to assess post-stroke dysarthria severity using multiple levels of speech representation. Notably, the computational approach employed here differs from commonly used expert feature methods. To achieve a syllable-level assessment, expert features in this study are computed at the syllable level rather than individually.

This paper gathered and organized existing acoustic expert features related to the four categories mentioned, resulting in the Phonation, Articulation, Prosody, and Intelligibility (PAPI) feature set. The dimensionality of the speech feature set $f$ was 20, capturing four components within the pronunciation process. **(1)** Phonation features included jitter, shimmer [47], Harmonics-to-noise ratio (HNR), Glottal-to-noise excitation ratio (GNE) [48], and Vocal fold excitation ratio (VFER) [49]. **(2)** Articulation included Jaw distance, Tongue distance, Movement degree, Vowel Space Area (VSA), Formant Centralization Ratio (FCR), Vowel Articulation Index (VAI) [50-52], the standard deviation of the first formant ($F_1$ std), the second formant ($F_2$ std), the third formant ($F_3$ std), and the intensity (Intensity std) features [53]. **(3)** Prosody included average intensity, syllable duration, and vowel duration features. **(4)** Intelligibility included vowel Goodness of Pronunciation (GOP) and consonant Goodness of Pronunciation (GOP) [54]. The acoustic features are shown in TABLE I.

TABLE I
ACOUSTIC FEATURES

| Type | Features |
|---|---|
| Phonation | Jitter, Shimmer, Harmonics-to-noise ratio (HNR), Glottal-to-noise excitation ratio (GNE), Vocal fold excitation ratio (VFER) |
| Articulation | Tongue distance, Jaw distance, Movement degree (F2i/F2u ratio), VSA, FCR, VAI, Variability of F1, F2, F3 and intensity |
| Prosody | Mean intensity, Syllable duration, Vowel duration |
| Intelligibility | Goodness of Pronunciation (GOP) of vowel and consonant |

In the visual analysis of this paper, expert knowledge was leveraged to design a lip-related feature set comprising amplitude, instability, and speed features of lip movements. Amplitude features included the minimum and maximum values of the average distance between the inner lips, as well as the minimum and maximum values of lip width during speech. Stability features encompass the stability of the left lip angle, right lip angle, and average inner lip distance during speech. Speed features included the velocity of left lip opening, right lip opening, and average inner lip opening during speech [24]. Consequently, a 10-dimensional set of lip features can be obtained for each vowel. The lip data is illustrated in Fig. 3, and the lip features are detailed in TABLE II.

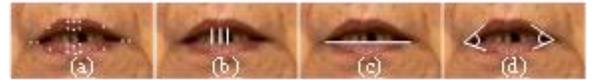

Fig. 3. (a) Lip landmarks, (b) Inner-lip distance, (c) Lip width, and (d) Lip angle.

TABLE II
VIDEO FEATURES

| Type | Features |
|---|---|
| Amplitude | Minimum internal lip distance<br>Maximum internal lip distance<br>Minimum lip width<br>Maximum lip width |
| Stability | Left lateral lip angle stability<br>Right lateral lip angle stability<br>Internal lip distance stability |
| Speed | Velocity of the left lateral lip angle<br>Velocity of the right lateral lip angle<br>Velocity of internal lip movement |

### C. Network structure

Patients with dysarthria exhibited characteristic features of vowels, including downward shifting in pronunciation, leading to compression or reduction of the vowel space. In both two-dimensional and three-dimensional vowel spaces, the vowel space of patients with dysarthria appeared more concentrated (see Fig 1). Based on this vowel space information, it has become a key representation in the field of dysarthria assessment [50-52]. However, existing methods only utilized low-dimensional vowel spaces for representation, neglecting the high-dimensional vowel space relationships, while more complete vowel space relationships offer better representational capabilities. Additionally, existing methods underestimated the importance of visual information for assessment. Therefore, the objective of modeling was to first integrate audiovisual information, and secondly, to explore more complete vowel space relationships. Based on this, a bi-modal audiovisual network is proposed.

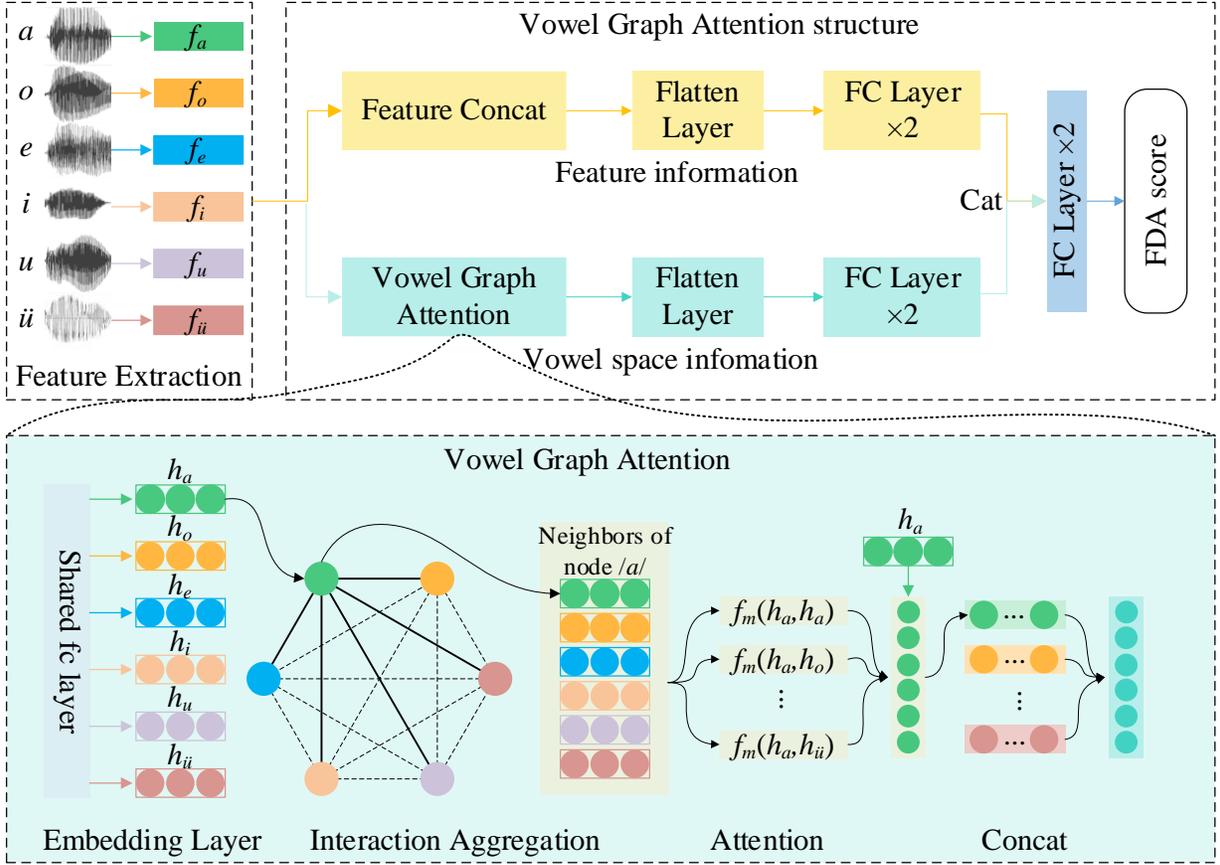

Fig. 4. Vowel Graph Attention Network structure. *f* is an acoustic expert feature, and the shared layer is a fully connected layer with a size of 16. The flatten layers in vowel graph attention are 576 and 120, respectively. The sizes of the two fully connected layers are (128,64). The fully connected layers after concatenation are (128,32).

The audio component included a Vowel Graph Attention (VGA) network and a Deep Neural Network (DNN). The VGA network utilized expert features of six cardinal vowels (/a/, /o/, /e/, /i/, /u/, /ü/) as graph node vectors, resulting in a graph with six nodes (representing the vowels) and a 20-dimensional feature for each node. Attention mechanisms established relationships (edges) between vowel nodes within VGAN, employing three attention heads with hidden layer nodes ($H=\{h_a, h_o, h_e, h_i, h_u, h_{ü}\}$) set to 16 per node. This structure was further processed through using two Dense layers (128 and 64 nodes, respectively) with ReLU activation for regression tasks. Simultaneously, the DNN component processed concatenated knowledge-based features through two fully connected layers (128, 64 nodes). The resulting output embedding layer was combined with the VGA embedding layer to form the acoustic representation. The VGAN network was specifically utilized to acquire speech vowel space information and expert feature information separately, which were subsequently fused to enrich the overall audio representation.

The vowel graph attention network structure is shown in Fig. 4. The input features, denoted as $F=\{f_a, f_o, f_e, f_i, f_u, f_{ü}\}$ were transformed into $H=\{h_a, h_o, h_e, h_i, h_u, h_{ü}\}$ after passing through a shared layer. Then, at the nodes, the matlul operation of $f_m$ was performed to calculate attention coefficients $e_{ij}$.

$$e_{ij} = f_m(Wh_i, Wh_j) \qquad (1)$$

wherein, $(i,j) \in \{a, o, e, i, u, ü\}$, $e_{ij}$ denotes the significance of the features of vowel $j$ to vowel $i$.

In this model, each vowel attended to every other vowel, including itself. To facilitate easy comparison of coefficients across different nodes, they were normalized using the *softmax* function over all choices for $j$. In the experiments, the attention mechanism *a* was a single-layer feedforward neural network parameterized by the weight vector *a* and utilized the LeakyReLU nonlinearity (with a negative input slope of α = 0.2). When fully expanded, the coefficients computed by the attention mechanism can be expressed as:

$$\alpha_{ij} = \frac{exp(\text{Leaky Re }LU(e_{ij}))}{\sum_{k \in (a,o,e,i,u,v)} exp(\text{Leaky Re }LU(e_{ik}))} \qquad (2)$$

The attention results between vowels were connected to obtain the vowel spatial relationships embedded in the input features, $\alpha$=concatenation$\{\alpha_{ij}\}$, $(i,j) \in \{a, o, e, i, u, ü\}$. It is worth noting that due to the use of fixed nodes in the network, batch parallel computation becomes possible, greatly improving computational efficiency. On the other hand, the input features were concatenated to obtain feature interrelationships embedded in $F$=concatenation$\{f_a, f_o, f_e, f_i, f_u, f_{ü}\}$. As shown in Fig 3, $\alpha$ and $F$ are each concatenated after two layers of fully connected layers to obtain the acoustic part embedding. The visual part embedding was formed by concatenating lip features, $V$ = concatenation$\{v_a, v_o, v_e, v_i, v_u, v_{ü}\}$. The final acoustic embedding and visual embedding were fused through cross-attention, as shown below:

$$E_{AV} = softmax((W_Q F)(W_K \alpha)^T) W_V \alpha \quad (3)$$

The regression model uses Mean Square Error, MSE) as the loss function, and its formula is as follows.

$$L_{MSE} = \frac{1}{m} \sum_{i=1}^{m} (y_i - f(x_i))^2 \quad (4)$$

where $m$ is sample numbers, $y$ is target FDA score and $f(x)$ is the predicted FDA score.

On the other hand, in the video section, visual expert features were employed as input, followed by the connection of three fully connected layers (128, 64, 32). The resulting embedding layer serves as the visual representation. Finally, the acoustic part representation and the visual part representation were fused through cross-attention and connected to a fully connected layer of size 32 for the regression of FDA scores, as illustrated in Fig. 5.

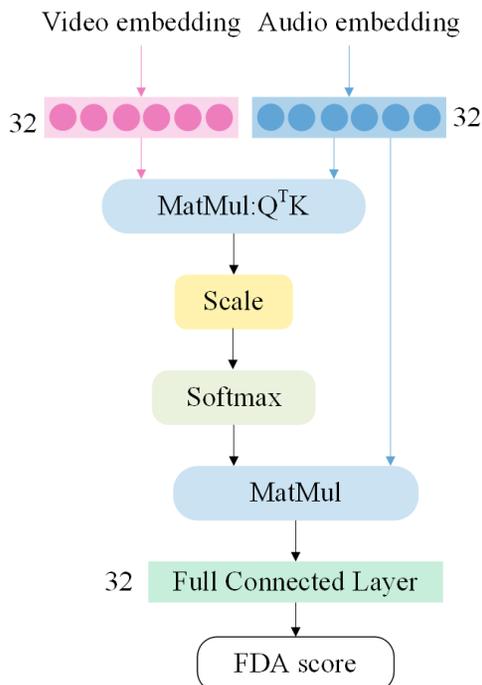

Fig. 5. Cross-attention between acoustic embedding and visual embedding

## IV. EXPERIMENTS

### A. MSDM Database

The MSDM [24] database is currently the largest Chinese dysarthria database, comprising audio and video data from 87 participants with post-stroke dysarthria, as shown in TABLE III. This group includes 62 subacute stroke patients with dysarthria (45 males, 17 females) and 25 healthy adults (13 males, 12 females). All patients underwent comprehensive clinical assessments provided by professional Speech and Language Therapists (SLTs) and neurologists. Cognitive abilities were assessed using the Montreal Cognitive Assessment (MoCA) [55], considering their cognitive decline history and mental status examination. The severity of dysarthria was evaluated using the Frenchay Dysarthria Assessment (FDA) [3]. The database covers a range of PSD patients, from mild to severe, with FDA scores ranging from 37 (severe) to 116 (mild). MSDM, designed based on the features and structure of the Chinese writing system, incorporates a series of stimuli to examine the quality of dysarthric speech. These stimuli include syllables, characters, words, sentences, and spontaneous speech tasks.

TABLE III
THE DURATIONS AND NUMBER OF PARTICIPANTS IN FIVE TASKS AND FOUR SEVERITY LEVELS

|  | Normal | Mild | Moderate | Severe |
|---|---|---|---|---|
| FDA scores | 116 | 87-116 | 58-86 | 37-57 |
| Participant Num. | 25 | 40 | 13 | 9 |
| Task | Duration (hours) | | | |
| Syllable | 2.77 | 1.35 | 0.89 | 0.47 |
| Character | 1.14 | 0.58 | 0.38 | 0.19 |
| Phrases | 1.44 | 1.54 | 0.95 | 0.63 |
| Sentence | 0.30 | 0.44 | 0.27 | 0.21 |
| Spontaneous | 0.42 | 0.90 | 0.33 | 0.15 |
| total | 6.07 | 4.81 | 2.82 | 1.65 |
|  | 15.35 | | | |

**Speech data processing**: The recorded audio files were segmented and transcribed using TextGrid in Praat. Syllable-level annotations were performed for each audio file. In this study, GMM (Gaussian Mixture Model) was employed to detect vowel segments in the pronunciation, with 70 mixture components, each having its general covariance matrix, and a maximum iteration of 60. **Video data processing**: RetinaFace [56] tracker was used adopted to detect faces, and the Facial Alignment Network (FAN) [57] was employed for landmark detection. Face alignment was performed by registering the face to the average face in the training set to eliminate size and rotation differences. The mouth's region of interest was cropped using an $80 \times 80$ bounding box. Each frame was standardized by subtracting the mean and dividing by the standard deviation of the training set. Finally, the three-channel color image was converted to grayscale through binarization. The resulting lip grayscale video served as input data.

### B Experimental setup

The experimental data utilized the aforementioned sample space construction method to augment the dataset, resulting in a sample quantity 17 times larger than the original (a total of 1,778,036 samples). The augmentation process ensured a balanced number of samples for each severity level. The dysarthria severity levels of participants in the training and testing sets were both balanced and non-overlapping. The ratio between the training set and the testing set was 9:1, with the use of 10-fold cross-validation. To evaluate the regression performance of the proposed approach, the Root Mean Square

Error (RMSE) and R-squared (R2) methods were employed.

## V. RESULTS

### A. Comparison of acoustic features

To evaluate the performance of the proposed PAPI acoustic expert feature set, PAPI was compared against commonly used acoustic features for dysarthria including MFCC, eGEMAPS, and GOP. The mentioned MFCC features consisted of 13-dimensional static coefficients extracted using the librosa library[58]. The eGEMAPS features comprised 88-dimensional acoustic features calculated by opensmile[59]. GOP utilized the GOP tool in kaldi[60], with a model consisting of a 13-layer tdnn trained on multi_cn data (approximately 2300 hours). The results indicated that the proposed acoustic feature set achieved superior performance compared to existing features, as shown in TABLE IV. In the regression of FDA scores (ranging from 0 to 116) on the VGAN network, the PAPI feature set performed best, with an RMSE of 11.89 and an R2 of 0.74 at the subject level. Compared to GOP, the RMSE decreased by 22.1% at the subject level. In comparison to eGEMAPS, the RMSE decreased by 21%. Furthermore, the eGEMAPS still demonstrated effectiveness when compared to traditional features, highlighting the advantage of VGAN in extracting vowel space information. This advantage is supported by the fact that eGEMAPS included resonance-related features, confirming the superiority of VGAN in extracting vowel space information.

TABLE IV
COMPARISON OF ACOUSTIC FEATURES

| Feature | RMSE | R2 |
|---|---|---|
| eGEMAPS | 14.66 | 0.60 |
| MFCC | 21.74 | 0.14 |
| GOP[41] | 15.27 | 0.57 |
| **PAPI** | **11.89** | **0.74** |

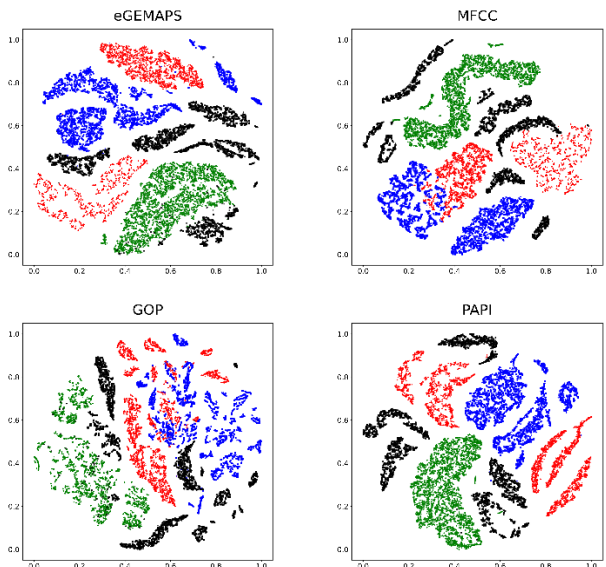

Fig. 6. t-SNE visualization demonstrates the contrasting effects of features. Different colors represent different categories.

As an attempt to further contrast the differences in dysarthria among different features, this experiment utilized t-SNE to visualize the penultimate embedding layer of the VGAN network, as shown in Fig. 6. Here, different colors were used to represented different participant groups, with green representing normal individuals (FDA scores 116 and above), black representing mild dysarthria (FDA scores 87-115), red representing moderate dysarthria (FDA scores 58-86), and blue representing severe dysarthria (FDA scores 37-57). From Fig. 6, it is evident that the embedding layer of PAPI exhibits a more pronounced differentiation compared to other features.

### B. Classifier comparison

In situations where data is limited, traditional machine learning methods sometimes achieve better performance than shallow neural networks. This is also the reason why early methods for the objective assessment of dysarthria often relied on traditional machine learning approaches. In this project, VGAN with commonly used traditional regressors (ADBOOST, GDBT, RF, SVR) on the PAPI feature set to assess the performance of the proposed VGAN network was proposed. The results indicated that the proposed network could achieve better results than traditional methods, as shown in TABLE Ⅴ. For regression on FDA scores (0-116 points) using the PAPI feature set on the VGAN network, the best performance was observed, with RMSE and R2 showing a 19.69% decrease and a 21% decrease compared to GDBT. When compared to SVR, there was a 31.43% decrease in RMSE and a 28.63% decrease.

TABLE Ⅴ
CLASSIFIER COMPARISON

| Model | RMSE | R2 |
|---|---|---|
| ADBOOST | 16.10 | 0.54 |
| GDBT | 15.06 | 0.59 |
| RF | 17.20 | 0.47 |
| SVR | 16.66 | 0.51 |
| **VGAN** | **11.89** | **0.74** |

### C. model comparison

In TABLE VI, VGAN was compared with the latest methods. In [26], a 6-layer CNN structure and MFCC features were used, while [40] employs a RESNET structure, STFT features. [41] and WAV2VEC2.0 [25], [61] used a 3-layer MLP. The results showed that the proposed method achieved optimal performance. For regression on FDA scores (0-116 points) using the PAPI feature set on the VGAN network, the best performance was observed with RMSE 11.89, and R2 0.74. Compared to MLP[41], there was a 21.15% decrease in RMSE. The results indicated that the proposed features and network architecture can more effectively extract representations related to dysarthria. Comparative results between wav2vec and wav2vec with augmentation data revealed that the vowel construction method mentioned in this paper exhibits higher stability.

Additionally, rows 8, 9, and 10 in Table VI display the results of ablation experiments. The VGAN consists of DNN and VGA

components, and the results show that compared to DNN, there is a 11.79% decrease in RMSE. Compared to VGA, there is a 19.66% decrease in RMSE, demonstrating the complementary nature of information between feature interaction and vowel space information.

TABLE VI
METHOD COMPARISON

| Model | Feature | RMSE | R2 |
|---|---|---|---|
| CNN[26] | MFCC | 18.85 | 0.34 |
| RESNET[27] | STFT | 18.76 | 0.08 |
| MLP[41] | GOP | 15.08 | 0.58 |
| DNN | WAV2VEC2.0 | 19.72 | 0.24 |
| DNN | WAV2VEC2.0(Augment) | 17.89 | 0.43 |
| DNN | HUBERT | 19.87 | 0.23 |
| DNN | PAPI | 13.48 | 0.66 |
| **VGA** | **PAPI** | **14.80** | **0.59** |
| **VGAN** | **PAPI** | **11.89** | **0.74** |

*D. Audio-visual modal fusion*

This experiment aims to explore the complementarity of audio and video modalities in various FDA tasks. Fig. 7 presents a comparison of the RMSE for acoustic modal, visual modal, and audio-visual bi-modal models across eight sub-projects (lips, reflex, jaw, laryngeal, respiration, velum, tongue, intelligibility). The FDA scores range for each sub-project is as follows: (lips 0-20, reflex 0-12, jaw 0-8, laryngeal 0-16, respiration 0-8, velum 0-12, tongue 0-24, intelligibility 0-16). To facilitate a more intuitive comparison of the RMSE results in the experiment, normalization was applied by dividing the RMSE scores of different sub-items by their respective maximum scores. It can be observed that, in the aspects of lips, jaw, tongue, and intelligibility, the bi-modal structure outperformed the auditory and visual modalities. Equivalent to the speech modality, the bi-modal RMSE showed a decrease of 15% for lip, 51% for jaw, and 13% for respiration, with a corresponding 39% decrease in intelligibility. This reflects the complementary nature of the visual modality to acoustics in this context. In fact, this result is logical because visual data only provides information about lips, jaw, respiration, and speech.

In addition, as shown in TABLE VII, the lowest RMSE of 11.12 can be achieved by combining auditory and visual modalities. Compared to the pure audio modality, the RMSE parameter of the bi-modal approach decreases by 1.38%. Similarly, compared to the video-only modality, the RMSE parameter decreases by 43.35%. These findings suggest that combining speech and visual modalities leads to a substantial enhancement in capturing information related to the speech quality of individuals with pronunciation disorders, thereby significantly improving the model's performance and enhancing its robustness. Additionally, this experiment demonstrates that the fusion method utilizing cross-attention is more effective than the concatenation fusion method.

TABLE VII
AUDIO-VISUAL MODALITY COMPARISON

| Modarity | Feature | RMSE | R2 |
|---|---|---|---|
| **Visual** | LIPFeature | 21.77 | 0.31 |
|  | CNN_Transformer | 19.72 | 0.29 |
| **Audio-visuao** | PAPI+LipFeature(cat) | 11.43 | 0.78 |
|  | PAPI+ CNN_Transformer | 11.43 | 0.76 |
|  | **PAPI+LipFeature(crossAtt)** | **11.12** | **0.78** |

VI. DISCUSSION

Vowels constitute a core component of syllables and they contribute significantly to speech intelligibility. Acoustically, the resonance patterns of vowels reflect the resonant features and frequency responses of vocal tracts configuration corresponding to each individual vowel. In general, the first formant (F1) corresponds to the height of the tongue during pronunciation, while the second formant (F2) to the frontness of the tongue. Articulatory mechanisms related to vowel defects in dysarthric speech involve deviations and reduced velocity in the movements of the articulators including the tongue, lips, and lower jaw, as well as abnormal movement durations [62]. Vowel distortion can be considered a hallmark feature of dysarthria [63]. Typically, patients with dysarthria often exhibit problems in vowel production such as undershooting (i.e., the vowels fail to reach the normative resonance frequencies), leading to a compressed or reduced working vowel space [64]. The acoustic consequences of these deficits in vowel production have been extensively studied [10-12] and summarized by Kent et al. [65], including concentrated formant frequencies, reduced vowel space area (i.e., average vowel space), and abnormal formant frequencies for high vowels and front vowels. Other features included the instability of vowel resonance patterns and decreased F2 slopes [10]. To qualitatively assess the clarity of vowel production, various vowel-related features are used, such as Vowel Space Area (VSA), Vowel Articulation Index (VAI), Formant Concentration Ratio (FCR), and the second formant ratio of /i/ and /u/ (F2i/F2u) [66]. These features are calculated from the formants of vowels /a/, /i/, and /u/, which are the most commonly used vowels in human language and represent the extreme positions of a speaker's pronunciation space [67]. These acoustic metrics have been utilized by speech-language pathologists (SLPs) [23] to investigate speech development, vowel recognition, and speaker features in dysarthria such as apraxia of speech [32], cerebral palsy-related dysarthria [33], or Parkinson's disease (PD) related dysarthria [34].

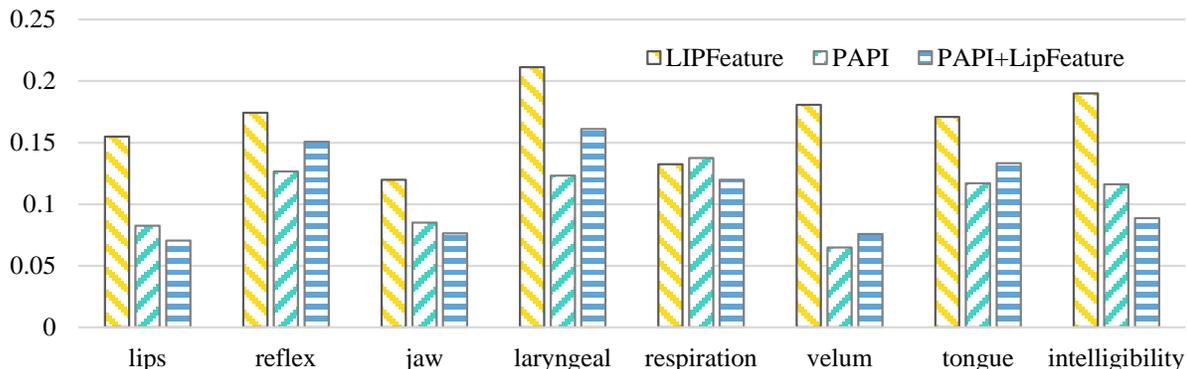

Fig. 7. Audio-visual modalities RMSE comparison in eight tasks

## VII. Conclusion

A novel modeling approach based on attention network to vowel space diagrams is proposed for dysarthria assessment. Firstly, this framework incorporated a set of more interpretable acoustic features specifically tailored for dysarthria assessment. Secondly, leveraging the vowel space theory, the Vowel Graph Attention Network (VGAN) used to extract spatial correlations between vowels using graph neural networks was introduced. Lastly, to enhance system robustness, visual information was integrated into the model. This method is found to be capable of accurately evaluating dysarthria data even with limited samples.

The proposed model was evaluated on the Mandarin Subacute Stroke Dysarthria Multimodal (MSDM) dataset for the regression task of overall FDA score and sub-category scores. Results demonstrated that the proposed method significantly improved dysarthria assessment accuracy, with audio-visual information showcasing complementary characteristics. In the future, exploration of correlations between different phoneme features and various FDA subclasses to establish more interpretable evaluation outcomes will be examined.

## Acknowledgment

This work was supported in part by the National Natural Science Foundation of China (U23B2018), National Natural Science Foundation of China (NSFC 62271477), Shenzhen Science and Technology Program (JCYJ20220818101411025), Shenzhen Science and Technology Program (JCYJ20220818102800001), Shenzhen Peacock Team Project (KQTD20200820113106007).